# Temperature dependent giant resistance anomaly in LaAlO$_3$/SrTiO$_3$ nanostructures

M. Z. Minhas[1], A. Müller[1], F. Heyroth[2], H. H. Blaschek[1], and G. Schmidt*[1,2]

[1]*Institut für Physik, Martin-Luther-Universität Halle-Wittenberg, Von-Danckelmann-Platz 3, D-06120 Halle, Germany*

[2]*Interdisziplinäres Zentrum für Materialwissenschaften, Martin-Luther-Universität Halle-Wittenberg, Heinrich-Damerow-Str. 4, 06120 Halle, Germany*

e-mail: georg.schmidt@physik.uni-halle.de.

**The resistance of the electron gas at the interface between the two band insulators LaAlO$_3$ (LAO) and SrTiO$_3$ (STO) typically drops monotonically with temperature and R/T curves during cooling and warm-up look identical for large area structures. Here we show that if the LAO/STO is laterally restricted by nanopatterning the resistance exhibits a temperature anomaly. Warming up nanostructures from low temperatures leads to one or two pronounced resistance peaks between 50 and 100 K not observed for larger dimensions. During cool-down current filaments emerge at the domain walls that form during the well-known structural phase transition of the STO substrate. During warm-up the reverse phase transition can interrupt filaments before the sheet conductivity which dominates at higher temperature is reestablished. Due to the limited number of filaments in a nanostructure this process can result in a complete loss of conductance. As a consequence of these findings the transport physics extracted from experiments in small and large area LAO/STO structures may need to be reconsidered.**

The formation of an electron gas at the interface between the two band insulators LaAlO$_3$ (LAO) and SrTiO$_3$ (STO) was discovered in 2004 [1]. Since then its origin has been under debate and the most prominent explanations are the so called polar catastrophe [1,2,3,4] or the presence of oxygen vacancies [5,6,7,8,9]. Up to now temperature dependent transport in this material system has mainly been investigated in large area structures. In these experiments the resistance typically drops monotonically upon cooling. During warm-up the resistance follows the same temperature dependence as during cooling down. This behavior is in agreement with both transport models mentioned above. Also a few results for transport in nanostructures are reported. In these examples different methods were used for patterning the 2DEG. In some cases conductivity was locally induced in a non-conducting interface using a conductive atomic force microscope (AFM) tip (for example [10]). In other cases the conductivity was modulated either by the patterning of an amorphous layer on the STO substrate prior to deposition of the crystalline LAO [11] or by low energy ion beam irradiation [12]. In all these cases, a monotonous drop in resistance with decreasing temperature was observed. Only recently an etching process was demonstrated by which the electron gas can be patterned into stable nanostructures while maintaining the interface conductivity and keeping the substrate insulating [13]. With these structures temperature dependent resistance measurements have been carried out which yield a surprising result. While the cooling curve corresponds to the one observed for large area structures the warm-up curve exhibits one or two massive peaks in resistance which typically occur at the temperatures associated with structural phase transitions of the STO substrate. We have performed a series of experiments on a number of different samples in order to identify the origin of the effect.

## Results

Fig.1 (a) shows a typical temperature dependence of the conductance of a large area Hall-bar. During cooldown the resistance decreases monotonically. During warmup the curve is reproduced except for a small hysteresis. This artifact is due to a lag of the sample temperature with respect to the sensor. Even when the sample has been kept at low temperature for days no deviation appears. For Hall-bars with a width of less than 500 nm the cooling curve is similar to the one of a large area structure. During warmup, however, a strong non-monotonicity is observed. At a certain temperature the resistance increases up to a maximum which can even be much higher than the resistance at room temperature (Figs. 1 and 2). When the temperature is further increased the resistance decreases again and finally the resistance curve again joins the one obtained during the cooling process. This resistance peak typically occurs at approx. 80K. In some cases this is followed by a second smaller peak well above 100 K.

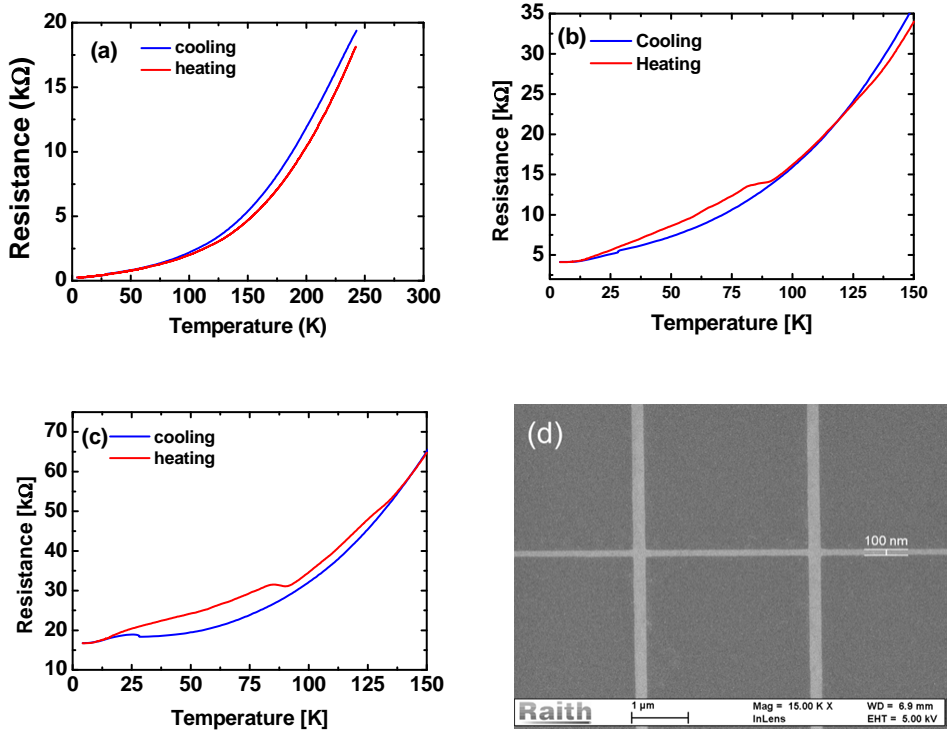

*Figure.1:* Temperature dependent resistance of the q2DEG in a large Hall bar structure (a), a 400 nm wide Hall bar (b), and in a 200 nm wide Hall bar (c). For the nano Hall-bars a non-monotonic behavior is visible with a resistance peak at approx. 80 K. For 400 nm width the resistance at 80K is only about 15% higher than in the cooling curve, for 200 nm the difference is 20%. (d) Shows an SEM picture of a 100 nm wide LAO/STO Hall-bar.

While we never observe this behavior for large area structures it is almost universal for nano-sized Hall-bars. Only the exact shape and height of the resistance peak varies from structure to structure. The effect is reproducible; however, the height of the peak depends on the history of measurements.

In order to identify the origin of the phenomenon we have done a series of investigations varying a number of parameters.

### Temperature dependence

In a first set of experiments we vary the minimum temperature ($T_{min}$) of the cooling cycles. The sample is always kept for one hour at $T_{min}$ before the sample is warmed up. As long as $T_{min}$ is below the peak temperature $T_{peak}$ the peak appears reproducibly at the same temperature of approx. 80K (see Fig.2), however, the peak height increases when the minimum temperature is decreased. When the minimum temperature approaches the temperature of the resistance peak the effect vanishes completely. The data is shown in Fig.2.

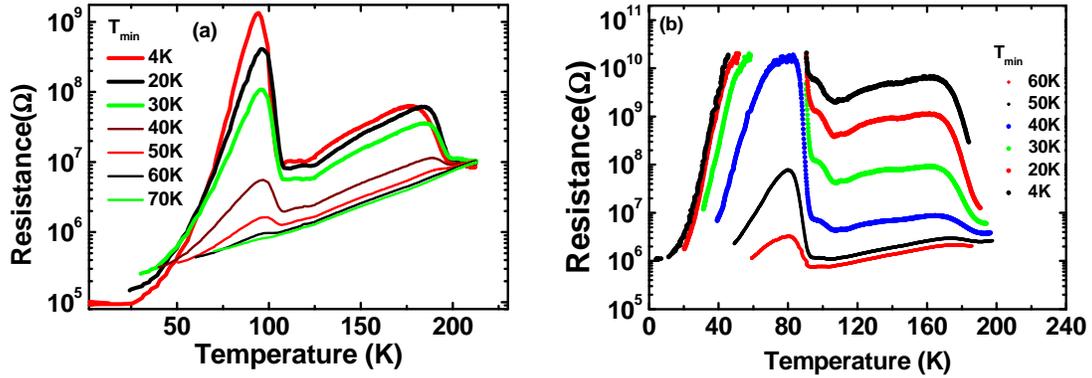

*Figure.2:* *Warm-up curves for two 100 nm wide Hall bars starting at different minimum temperatures $T_{min}$. For lower $T_{min}$ the peak is much higher. For the 2$^{nd}$ sample (b) the resistance maximum is well beyond the measurement limit for $T_{min}$ < 40 K.*

### Stability over time

As the time constants of the effect may give some insights into its physics we also investigate two aspects of time dependence. In a first series of experiments we vary the waiting time at the lowest temperature of 4.2 K. The minimum time is obtained by cooling down and immediately warming up when 4.2 K are reached, which at the cooling rates that we use corresponds to a few minutes at a temperature close to 4.2 K while the maximum waiting time is 24 hours at 4.2 K. In these experiments we do observe small variations (as is explained below the measurement history can influence the experiment) but no systematic change in peak height.

In a second experiment (Fig. 3) we stop the warm-up procedure for approx. 12 minutes at 60 K and 70 K (below the peak temperature), respectively, and a third time at 80 K which for this sample is the temperature of maximum resistance. The curves obtained at 60 K and at 70 K both show the following behavior. In both cases resistance has already increased. When warm-up is stopped the resistance increase continues for several minutes and finally saturates.

At 60 K the starting point of this curve is at $5 \times 10^6$ Ω and the resistance increases up to $2.5 \times 10^7$ Ω within 4 minutes. At 70 K the curve starts at $5 \times 10^7$ Ω and increases up to $2 \times 10^8$ Ω within another 4 minutes. At 80 K, however, the resistance does not increase any further but decreases massively after 10 minutes it has approximately reached the corresponding value of the cooling curve.

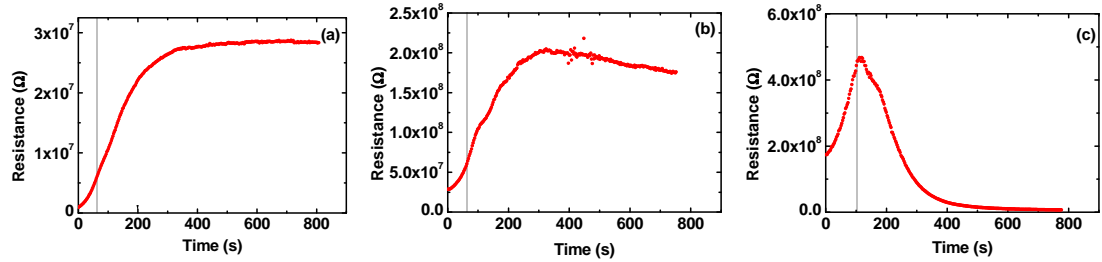

*Figure.3:* *Warm-up curves for a 100 nm wide Hall bar with waiting times at different temperatures. For (a) the sample is warmed from 50 K to 60 K (left hand side of grey line) and the temperature is then kept stable for approx.. 12 minutes (right hand side of grey line). (b) Similar as (a) only now the sample is heated from 60 K to 70 K where the temperature is kept stable. In (c) the process is repeated now heating from 70 K to 80 K. While below the peak maximum (a and b) the resistance further increases while the temperature is constant beyond the peak maximum at 80 K the resistance drops over time although the temperature is constant.*

## Sample type

As mentioned above samples have been fabricated using a subtractive etching process. Although already showed that the etch damage only influences a narrow region at the side of the nano-structures we want to make sure that the observed effect is not an artifact related to the patterning process. In order to confirm the basic nature of the observation we test additional samples fabricated using a different process introduced by Schneider et al [14]. Also for these samples we see a pronounced peak as shown in Fig. 4 confirming that the lateral size restriction is the relevant factor for the increase in resistance.

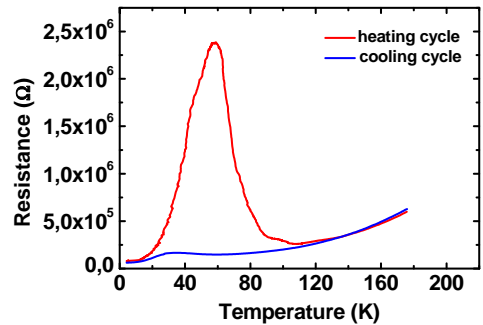

*Figure.4:* *Cooling and warm-up curve for a 100 nm wide Hall bar fabricated using the process described by Schneider et al [14]. Again a massive peak is observed. Even a small resistance peak is visible in the cooling curve.*

## Discussion

From the temperatures at which the resistance peaks occur it is likely that the effect originates from the various structural phase transitions that STO undergoes during cooldown and warmup. These phase transitions lead to the formation of structural domains in the material and are known to occur at approx. 110 K (from cubic to tetragonal) and 65 K (from tetragonal to orthorhombic)[15]. The upper phase transition temperature will be labelled $T_{C1}$ in the following while the lower one is $T_{C2}$. A third transition to a rombohedral phase well below 30 K is also suspected [15]. In 2013 Kalisky et al. [16] showed that at low temperature the current distribution in large area LAO/STO heterostructures is no longer homogeneous but an increased current density exists along the domain boundaries in the STO substrate. In these experiments a filamentary pattern of higher current density was observed at T= 4.2 K by scanning SQUID microscopy in agreement with a typical domain pattern in the STO. The experiments, however, yielded no information on the magnitude of the conductivity of the domains themselves. At the same time Honig et al. [17] identified a large anomalous piezoelectricity with a strong local variation related to the domain boundaries in the STO substrate. In 2016 [16] was succeeded by a second publication in which more detailed measurements of a similar type were presented [18]. In this work it was shown that due to the domain structure the conductivity is modulated by at least 95 %.

This filamentary current distribution alone, however, is not sufficient to explain the non-monotonic temperature dependence that we observe in nanostructures. Especially as the latter only appears during warm-up and not during cool-down additional physics needs to be taken into account:

When during cool-down temperature falls below $T_{C2}$ (Fig. 5a) the appearing domain walls exhibit large electric fields due to the piezoelectricity and the large dielectric constant. Charge and thus current flow start to accumulate at the domain boundaries as described in [16] (Cooling just below $T_{C1}$ yields no effect). This may be accompanied by the pinning of defects at the domain walls. These defects can trap and release carriers and lead to longer time constants for charging and discharging. The fact that the dielectric constant of STO is massively increased at lower temperatures may additionally stabilize the current paths at lower temperatures.

The originally homogeneous conductivity in the area between the domain walls is more and more reduced while filament conductivity increases, resulting in a maze of a limited number of conducting filaments at the domain boundaries and high-resistivity (Fig. 5b) or even insulating regions (Fig. 5c) inside the domains. This corresponds to the state observed in [16]. Due to the way this maze is formed below the phase transition temperature there are always conducting paths. During cooling this is of no visible consequence because the total amount of charge does not necessarily change and the mobility can increase anyway leading to the typical decrease in resistance with decreasing temperature. In fact the accumulation of charge can even lead to additional screening which further reduces scattering. The effect itself is reminiscent of conducting domain walls in insulating $BiFeO_3$. Similarly in our samples large electric fields occur at the domain walls which can attract charges and charged defects that lead to our observations as described below.

When the sample is warmed-up, the situation is different because the current is already concentrated in filaments when the phase transition temperature is approached. When the temperature is further increased not all filaments break at once. On the contrary, at some point close to $T_{C2}$ a few domain boundaries disappear while the rest of the domain structure is still stable. As a result only a few filaments become unstable meaning that the charge is no longer pinned by local fields. The adjacent domain walls can now attract the remaining charge which was formerly pinned and thus increase the local resistance up to infinity. In large area samples this is of no consequence. With literally millions of parallel current paths, removing even thousands of them does not lead to a measurable increase in resistance. In a nanostructure, however, the current is limited to one or at least few filaments at low temperatures (Fig. 5d).

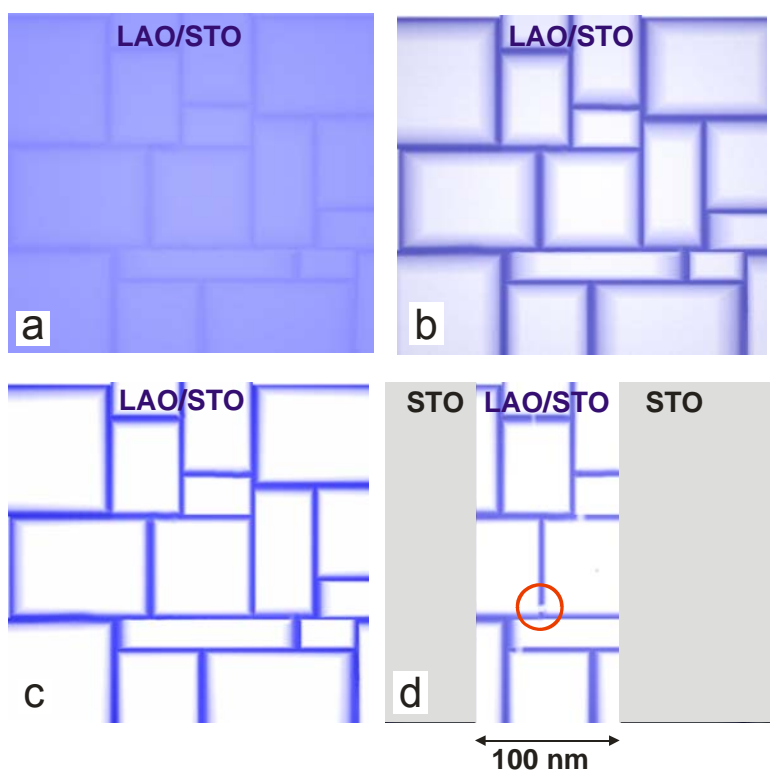

*Fig. 5: Simple sketch of a toy domain configuration which can cause the effect observed in our experiments. Dark Blue (gray) regions correspond to higher conductivity, light areas have low conductivity, white is insulating. (a) Below $T_{C2}$ filaments form and oxygen vacancies start to accumulate at the domain boundaries. (b) After some time the domain boundaries are the main conducting paths, the inside of the domains is slightly conducting. (c) Finally the full conductance of the sample is through the domain walls. Areas are insulating. (d) Situation in an etched nanostructure at warm-up. The STO (sides) is insulating while in the center (LAO/STO) filaments persist. The domain walls have disappeared; however, the vacancies have not yet been redistributed. If at a critical position a filament breaks (circle) no conductance is left.*

The breaking filament then leaves a system of lower conductance if an alternative current path exists. Starting from two filaments breaking a single one just doubles the resistance. If, however, originally conduction was through a single filament the resistance may increase dramatically if not become unmeasurable. At temperatures below the phase transition this situation is stable over time as charge is still bound to the domain boundaries and the broken link cannot be mended. Only when the temperature is raised beyond $T_{C2}$ all domain walls disappear charge becomes mobile again and is redistributed with a certain time constant. Still even above $T_{C2}$ this redistribution will take some time and not be instantaneous. Depending on the domain geometry the process may even be repeated at $T_{C1}$ if the charge is not yet fully redistributed leading to a second peak in resistance.

Obviously the total number of filaments is bigger in bigger structures. It should, however, be noted that no direct proportionality is expected because of the statistical distribution of lateral domain size. Already from a statistical point of view this picture nicely corresponds to our observations. We have not observed any peak in resistance for large area structures. Nanostructures with a width of less than 500 nm, however, always show some peak in resistance. For some of them, especially for the small ones (100 nm), the resistance increases above the room temperature value or even beyond the measurement limit. It should be noted that there are such observations by a number of groups [6,19,20] who have observed a monotonous drop in resistance during cool down and small non monotonicities in resistance during warmup in a large area sample, however, the authors could give no explanation. These results will be discussed later.

We now compare this model with our additional experimental data. The finite rise time of the resistance peak (Fig. 3) is explained by the nature of the phase transition during warm-up. The phase transition happens over a certain time and temperature range. Taking into account that also the lattice contracts further during cooling or expands during warm-up, the domain structure cannot be seen as static but as something which can always undergo small modifications when the temperature changes close to $T_C$. So the increase in resistance happens gradually over a certain temperature range and starts already below $T_C$. Also the charges need to migrate away from the vanishing domain wall towards the remaining ones which can be a slow process. This fits the rise in resistance when warm-up is stopped at 60 K or 70 K (Fig. 3 a and b) which apparently is below the phase transition in this sample. At 80 K which is above the phase transition the domain walls vanish and charges become mobile leading to a slow drop in resistance over time (Fig. 3 c). The time constant of this decrease is determined by charge diffusion and by detrapping of defects because there is no other restoring force to redistribute the charge in the sample. Depending on the domains structure the size of the rupture and the geometry of electric fields are different and we expect a variation of peak shape and stability over time not only from sample to sample but in some cases even from measurement to measurement.

It is important to realize that the maximum resistance is determined by the degree of depletion of the areas between filaments and not by the high conductivity of the filaments themselves. A simple gedankenexperiment shows that if the sheet conductance were reduced to 1% of the original value the resulting increase in resistance for all filaments breaking still cannot be more than two orders of magnitude. The increase observed in our case indicates quasi fully insulating domains. With this in mind the model also fits the experiments with different $T_{min}$. Cooling to lower temperatures takes more time and leads to stronger accumulation at the domain boundaries especially because the dielectric constant of the STO increases at lower temperatures. This leaves less charge carriers for the sheet conductance at lower T. As the sheet conductance determines the peak height the peak must be higher for lower $T_{min}$ as seen in the experiment even if the pattern of filaments is identical in all cases. The second peak which is sometimes observed (Fig. 2 a) is also readily explained. When the resistance has decreased from the peak maximum above to the original value above $T_{C2}$ this sheet conductivity is not necessarily already completely homogeneous. It only indicates that a new maze is established. The next phase transition at higher temperatures can thus again interrupt the current flow, however, only if the temperature had fallen below $T_{C2}$ during the experiment.

It is not even self-evident that at room temperature the carriers again reach a completely homogeneous distribution. While at low temperatures the electric fields at the domain boundaries can collect the charge carriers there is no electric field except for possible Coulomb repulsion which reestablishes a uniform distribution when the domain walls are no longer present. As a consequence, any further experiment can be influenced by the history of measurements and the peak height may vary even for apparently identical temperature cycles. The results are also consistent with the AFM induced conductivity mentioned in the beginning where electric fields create local conducting paths in an insulating environment which are stable up to room temperature.

Theoretically nanostructures may exhibit a peak in resistance even during cooldown. When the filaments are established below 80 K the third phase transition below 30 K can theoretically cause a similar scenario. Indeed for at least one structure we observe a resistance peak during cooldown below 30 K (Fig. 4) especially after several temperature cycles. Finally with this in mind also the observations in large area structures [6,19,20] can be explained. Although there the effect should normally not occur it is possible that the current distribution in a large area is inhomogeneous and close to the percolation limit. In this case, the disappearance of a few filaments during warm-up can at least slightly decrease the conductance as observed in [6,19,20]. However, the statistical probability for this effect to be observed is very low.

Although the picture outlined above yields a good explanation for our observation we would like to discuss a few alternatives. As the effect appears at the phase transition temperature it is likely to be related to the domain structure and the question arises whether conducting domains and insulating domain walls might also be a suitable scenario. This can be quickly dismissed as vanishing domain walls during warm up would lead to a decrease in resistance rather than an increase. Also it is important to consider the role of defects induced by the etching process which might also change the temperature dependence of the resistance. Here a number of arguments can be setup against. Indeed a depletion region had been observed in [13]. This region, however, was very narrow and located at the edges of the structure. There might be an interplay of these defects with the domain boundaries, however, we would expect for a strong decrease of the effect when the width is increased from 100 nm to 200 nm, which is not observed. Also we have used two different patterning processes, one even without etching and both lead to similarly large results. It thus seems that conducting domain walls with insulating regions in between are the most likely explanation, although we are not able to determine the exact pinning mechanism.

Finally two examples of nanopatterned LAO/STO structures should be discussed which at first glance seem to contradict our findings [11,12]. These experiments have investigated nanostructures which were not fabricated by conducting AFM and both experiments have yielded a completely monotonous dependence of resistance on temperature. There is, however, a likely explanation. Firstly Stornaiulo et al. [11] have only investigated a minimum width of 500 nm which is above the threshold of 400 nm that we observe. Aurino et al. [12] however, have used a minimum structure of 100 nm. Nevertheless, none of the two papers compares cool-down and warm-up curves and most likely the measurements shown were done during coold-down.

## Summary

We have shown that LAO/STO nanostructures show one or two peaks in resistance when warmed up from low temperatures. The peaks occur at the temperatures of structural phase transitions of the STO. The peak height depends on the minimum temperature of the cooldown. Above the critical temperature the maximum resistance is not stable over time but goes back to minimum after approx. 10 minutes. We can explain the behavior by the formation of current filaments at the boundaries of the domains which appear during the structural phase transition at $T_{C2}$ in the STO and which may break during warm up. During cool down the formation starts from a homogeneously conducting area and is mainly undetectable. During warmup a breaking filament can disrupt localized current paths in a nanostructure leading to the observed increase in resistance. The filaments are formed because large electric fields at the domain boundaries together with an increasing dielectric constant lead to an accumulation of charge carriers at the boundaries leaving area inside the domain basically insulating. While the effect is only observed during warm-up through the phase transition it is also of importance for the interpretation of any transport experiment in LAO/STO below $T_{C2}$ because the assumption of a homogeneously conducting interface seems to be no longer valid.

## Methods

### Growth

All films used in our experiments are deposited by pulsed laser deposition (PLD) as described previously [13]. During growth the background oxygen pressure is $10^{-3}$ mbar. LAO layers are deposited from a single crystal LAO target on $TiO_2$-terminated STO (001) substrates [21,22]. The substrate temperature during deposition is 850°C. Laser fluence and pulse frequency are kept at 2 J/cm$^2$ and 2 Hz, respectively, during the deposition. Reflection high-energy electron diffraction (RHEED) is used to monitor the layer thickness with unit cell resolution during the growth. After deposition of 6 unit cells of LAO the sample is slowly cooled down to room temperature while the oxygen pressure is maintained. As a result we obtain layers with a sheet resistance of 133 Ω/□ and a typical mobility of $1.124 \times 10^3$ $cm^2V^{-1}s^{-1}$ at 4.2 K.

### Nanopatterning

These layers are then patterned into nanosized Hall-bars. For this purpose two alternative processes are used in order to allow us to check for possible artifacts from processing. In the first process a resist is patterned using negative electron beam lithography. The resist is then used as an etch mask in a reactive ion etching process which removes the LAO down to the STO substrate. Details of the process have been described in [13]. In the second process the recipe of Schneider et al. [14] is used, however, with some modifications. An epitaxial 2.u.c of LAO is grown on STO using the same parameters as described above. Subsequently positive e-beam lithography is done and amorphous LAO is deposited and patterned by lift-off. Then a 4.u.c layer of epitaxial LAO is deposited in an oxygen pressure of $10^{-3}$ mbar. Epitaxial growth, however, can only occur in the places where no amorphous LAO is present. As a consequence only in these areas a conducting LAO/STO interface can form. The sample is then cooled down to room temperature in 1000 mbar of $O_2$. The cool down includes a 1 h annealing step in oxygen at 600°C. All resulting patterned structures are stable at ambient conditions. The samples are bonded and electrical transport measurements are carried out in a $^4$He bath cryostat.

### Measurements

The experiments are carried out in a $^4$He bath cryostat with a variable temperature insert. The samples are cooled down at a rate of approx. 5 K/min and warm-up is done at a rate of approx. 2.5 K/min. The resistance is measured in a four probe geometry using a nanosized Hall-bar. The active region between the voltage leads has a length of 3.5 μm and the width which is indicated in the text. The applied DC voltage is 80 mV and the current is measured using a 1 MΩ series resistor. Voltages are measured using custom made zero drift voltage amplifiers and an Agilent 34420A 7.5 Digit nanovoltmeter.

## Acknowledgement

The work was supported by the European Commission in the project IFOX under grant agreement NMP3-LA-2010-246102 and by the DFG in the SFB 762.

## Author contributions

M.Z. Minhas performed most of the transport measurements and wrote the manuscript. A. Müller performed electron beam lithography and some transport measurements. F. Heyroth performed electron beam lithography. H.H. Blaschek processed the samples after lithography, G. Schmidt planned and supervised the experiment and wrote the manuscript.

## Additional information

The authors declare no competing financial interests.